\documentstyle[aps,epsf,pre,twocolumn]{revtex}

\draft 

\begin{document}

\title{Reversible Random Sequential Adsorption of Dimers on a Triangular
Lattice}

\author{R. S. Ghaskadvi and Michael Dennin}
\address{Department of Physics and Astronomy}
\address{University of California at Irvine}
\address{Irvine, CA 92697-4575.}

\date{\today}

\maketitle

\begin{abstract}

We report on simulations of reversible random sequential adsorption of
dimers on three different lattices: a one-dimensional lattice,
a two-dimensional triangular lattice, and a two-dimensional triangular
lattice with the nearest neighbors excluded. In addition to the adsorption
of particles at a rate $K^{+}$, we allow particles to leave the surface
at a
rate $K^{-}$. The results from the one-dimensional lattice model agree
with previous results for the continuous
parking lot model. In particular, the long-time behavior is dominated
by collective events involving two particles.
We were able
to directly confirm the importance of two-particle events
in the simple two-dimensional triangular lattice. For the
two-dimensional triangular
lattice with the nearest neighbors excluded, the observed
dynamics are consistent with this picture.
The two-dimensional simulations were motivated
by measurements of Ca$^{++}$ binding to Langmuir monolayers. 
The two cases were chosen to model
the effects of changing pH in the experimental system. 
\end{abstract}


\section{Introduction}

A large number of nonequilibrium systems can be qualitatively described
as a flux of particles impinging on a surface or line. Two heavily
studied models of such systems treat the particles as either
fixed in place upon impact (random sequential adsorption) or
as free to diffuse along the surface or line
(random cooperative adsorption) \cite{E93}. One can also
consider the deposition of particles that are free
to desorb \cite{SVS98,TST90,J94,JTW94}. Some
examples of the wide range of applicability of these
models include coating problems, chemisorption, physisorption,
the reaction of molecular species on surfaces and at
interfaces, and the binding of ligands on polymer chains.
Jamming is one of the common occurrences in these systems
that random sequential adsorption models effectively describe.
Loosely speaking, a jammed system is one that is locked
into a state of partial coverage
because of adsorbate size or shape.
In addition to the various adsorption processes, jamming
occurs in a wide range of nonequilibrium
situations, including glasses,
granular materials, and traffic flow \cite{RL87,MPV87,WSB96}.
In spite of significant progress, no general framework exists
for the description of jamming phenomena.

A particular realization of random sequential adsorption is
the parking lot model \cite{E93,R58,JTT94,PB97,KB94}.
In the irreversible version of this model, identical particles
(cars) adsorb on a line (curb) at a rate $K^{+}$. 
In this model, the phenomenon of jamming has been known
for some time \cite{R58}. A certain number
of the parked cars leave a space
that is too small to fit another car. These are
referred to as bad parkers. The result is a density of cars
along the curb that is less than one. The
density of cars reached in the irreversible model is the
jamming limit.

In the reversible
version, identical particles (cars) adsorb on a line (curb)
at a rate $K^{+}$ and leave the line (curb) at a rate $K^{-}$.
The removal of cars allows for adjustments in the bad parkers
that relieve the jamming.
Recently, there has been renewed
interest in the reversible case because of its successful
application to compaction in granular materials
when generalized to three dimensions \cite{NKBJN98}. In this
version, the ``parking spots''
are voids in the material that can be filled with particles.
The dynamics of the reversible parking lot model for large values
of $K = K^{+}/K^{-}$ has a number of interesting features. Perhaps
the most
dramatic feature is the existence of two very different time scales
for the evolution of the coverage fraction of particles \cite{KNT99}.
First, there is a rapid approach to a coverage fraction that is
equal to the jamming limit. This is followed by a slow relaxation
to a larger steady state value. The slow relaxation is
understood in terms of collective
parking/leaving events involving multiple cars \cite{KNT99}.

In this paper, we present the results for simulations of
the reversible adsorption of dimers on: (1)
a one-dimensional
lattice, (2) a two-dimensional triangular lattice,
and (3) a two-dimensional
triangular lattice with the nearest neighbors excluded.
The one-dimensional lattice model \cite{KB94,F39,CR63,W66}
was chosen as a test case, and the results are in good agreement
with existing data. In particular, our simulations
confirm the importance of collective parking events in controlling the
slow dynamics, as seen in Ref.~\cite{KNT99}. The
two triangular lattice models exhibit differences
in their time evolution that can be attributed to
effects of bond orientation and packing on the collective events.
The case without nearest-neighbor exclusion
corresponds to attempting to cover the plane with a shape
formed by two regular hexagons sharing a side.
The nearest-neighbor excluded case corresponds to a tiling
of distorted hexagons that cover multiple sites.

The rest of the paper is organized as follows. Section II describes
the details of the simulations. Section III presents the results
for the one-dimensional model. Section IV presents the results for
the two triangular lattices. The simulations
were motivated in part by experimental measurements of the
viscosity of Langmuir monolayers. A brief description of the
experimental system and its relationship to the simulations presented
here is given in Section V. The results are discussed and
summarized in Section VI.

\section{Simulation Details}

For the one-dimensional simulations, a line of 32000 particles was used.
Both of the triangular lattices consisted of a
grid of 1000 x 1000 particles.
To distinguish between the two-dimensional models, we
introduce the following nomenclature.
Model A will refer to the triangular lattice without nearest-neighbor
exclusion. Model B will refer to the
triangular lattice with nearest-neighbors
excluded from binding. Particles are taken to bind to two neighboring
sites on the lattice, forming a dimer. The binding occurs at a
rate $K^{+}$, and particles leave the surface at a rate of $K^{-}$.

At each step in the simulation, a site was chosen at random. Then, a
random number between 0 and 1 was compared with the ratio
$K^{+}/(K^{+} + K^{-})$ to
determine whether a binding or unbinding event was attempted.
For unbinding
events, if the chosen site was part of a bond, the bond was broken;
otherwise,
no action took place. For binding events, a nearest
neighbor was randomly selected. A binding event occurred
only if both sites
were allowed binding sites. The definition of allowed binding site
depends
on the model. For the one-dimensional and Model A cases, an allowed
site is any site that is not part of a bond. For
Model B, if either site is part of a bond or the nearest neighbor of
a bound site, binding is not allowed. It is important to note that the
number of new bonds created is directly proportional to the number of
allowed sites, which is not the same as the number of open
sites. The number of desorption events is still directly proportional
to the coverage fraction. The coverage fraction, $\rho$, is
defined as the ratio of
sites that are part of a bond to the
total number of sites.

A schematic of each of the model systems with examples of bound sites
is shown in Fig. 1. It is important to notice the different spatial
structures in Model A and Model B. In Model A, complete coverage
corresponds to all sites being part of a bond. In Model B,
perfect coverage of the system corresponds to a tile of distorted
hexagons that are composed of both empty and bound sites.
This results in a maximal coverage fraction
$\rho_{max} = 0.4$. For both the one-dimensional case and Model A,
$\rho_{max} = 1.0$. In this paper, $\rho(\infty)$ will
designate the steady
state value of the fractional coverage, and $\rho_{jam}$ will refer
to $\rho(\infty)$ in the case $K^{-} = 0$, i.e. the jamming limit.

\begin{figure}
\epsfxsize = 3.2in
\centerline{\epsffile{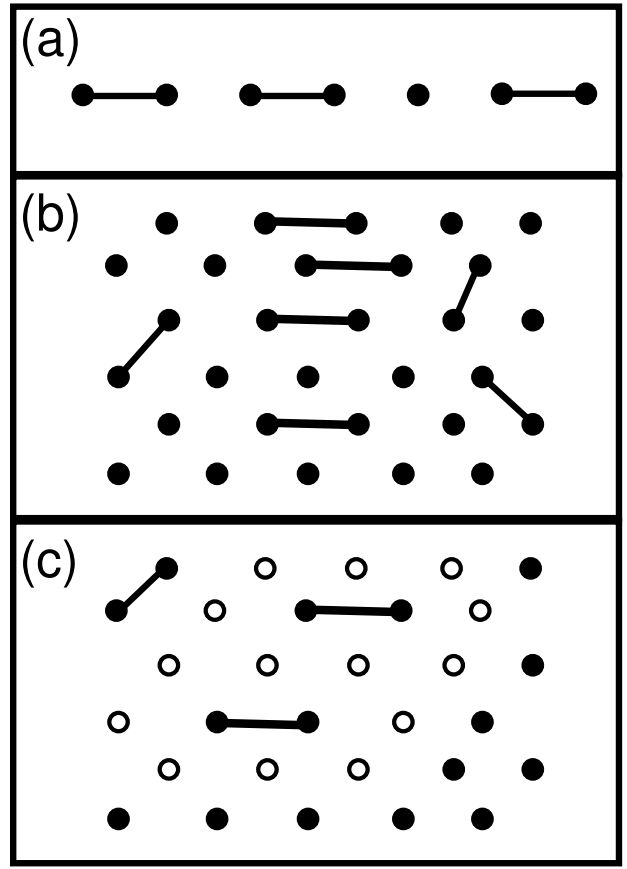}}
\caption{(a) Example of bonds in the one-dimensional lattice model.
The sites are represented by the dots, and the incoming particles are
represented by solid lines. The particles form
a bond between two neighboring sites.
(b) Example of bonds in Model A. In this case, particles can bind
any two nearest-neighbor sites that are not already part of a bond.
(c) Example of bonds in Model B. In this case, nearest-neighbors of
a bound site are not allowed to form bonds. Examples of such sites
are represented by open circles.}
\end{figure}

\section{One-dimensional Simulation}

Figure 2 shows $\rho$ as a function of iteration step for selected
values of
$K = K^{+}/K^{-}$.
We include the case $K^{-} = 0$, which gives $\rho_{jam} = 0.86474$. For
comparison, analytic calculations give
$\rho_{jam} = 0.86466$ \cite{F39,W66}.
The original work on the reversible parking lot model \cite{KB94}
proposed a mean-field description of the dynamics that can be
expressed in terms of the average density, or fractional surface
coverage, $\rho$. Both the continuous and lattice versions of the
parking lot model were considered. 
Figure 3 shows the steady state value of $\rho$ for the values of
$K$ plotted in Fig. 2.
The solid curve in Fig. 3 is the value for $\rho(\infty)$
for dimers binding to a one-dimensional lattice,
as determined by the following equation from Ref.~\cite{KB94}:
\begin{equation}
\rho(\infty) = 1 - (K^{-}/K^{+})^{1/2}/2.
\end{equation}
The agreement between our simulations and Eq. 1 confirms
the mean field prediction for the equilibrium values of
$\rho$. However, as with the continuous parking lot model \cite{KNT99},
the mean-field description is unable to accurately predict the
time evolution of $\rho$. This can be seen in Fig. 2 where the two
time scales controlling the evolution of $\rho$ are evident for $K > 10$.
The system rapidly reaches $\rho_{jam}$, and then slowly approaches
its equilibrium value. As K goes to infinity, $\rho(\infty)$ approaches one,
but the time to reach equilibrium approaches infinity. This is in
agreement with results for the continuous parking lot model reported in
Ref.~\cite{KNT99}.

\begin{figure}[htb]
\epsfxsize = 3.2in
\centerline{\epsffile{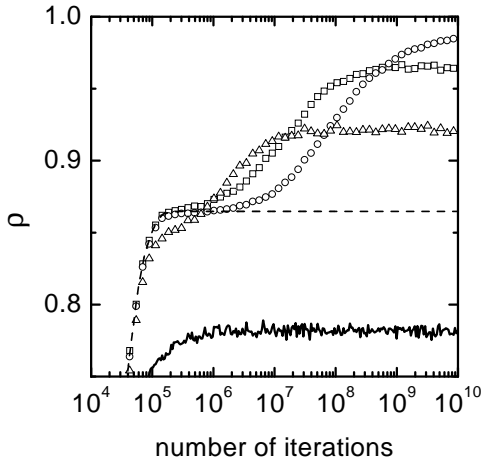}}
\caption{Shown here is the coverage fraction $\rho$ as a function of
the number of iterations for the one-dimensional model. The dashed
line represents the evolution for $K^{-} = 0$. The other curves are
for: solid line ($K = 5$), open triangles ($K = 40$),
open squares ($K = 200$), and open circles ($K = 1000$).}
\end{figure}

We have found that the explanation
of the two distinct time scales reported in
Ref.~\cite{KNT99} applies to the discrete case as well.
Essentially, collective events are responsible for the
evolution of $\rho$ for $\rho > \rho_{jam}$. In
Ref.~\cite{KNT99}, the authors calculated the transition
rates for two good particles to one bad particle
and one bad particle to two good particles and found
that these rates account for the additional slow
time scales. In contrast, we directly monitor the
transitions as part of the simulation. The reason
such transitions result in an additional slow time
scale can be understood in terms of the following argument.

\begin{figure}[hb]
\epsfxsize = 3.2in
\centerline{\epsffile{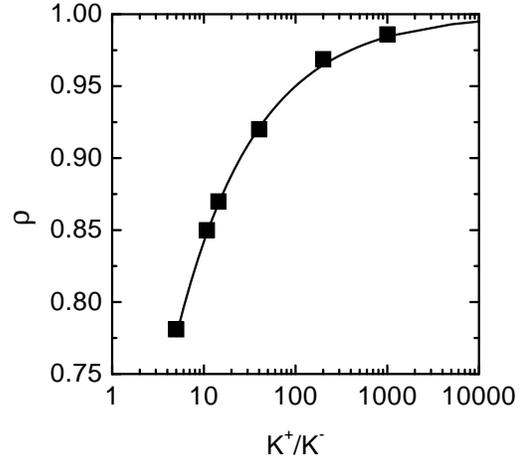}}
\caption{The solid line is the mean-field prediction for
$\rho(\infty)$ as a function of $K$. The symbols are the 
values of $\rho(\infty)$ taken from Fig. 2.}
\end{figure}

As discussed in the introduction, when $K^{-} = 0$, jamming
occurs because of ``bad parkers'' that leave empty space.
For the one-dimensional lattice, empty space refers to a
single site that is unable to bond. An example is shown in
Fig. 4a. For small values of $K^{-}$, bad parkers initially occur
at essentially the same rate as for $K^{-} = 0$
because very few particles desorb. Therefore,
the coverage fraction for the system quickly approaches a
value of $\rho_{jam}$.
Even when a value of $\rho_{jam}$ is reached,
the rare desorption event is generally
followed immediately by a readsorption because $K^{+}$ is so large.
The total number of particles is not changed
by these events. However, when one bad parker desorbs
and two particles adsorb in the opened good locations,
then the number of particles is increased by one. Likewise,
if two good parkers unbind and one bad parker binds, the number
of particles is decreased by one. Because these events involve
multiple particle transitions, they occur on a longer time scale
then simple adsorption/desorption events.

For the one-dimensional discrete case, one can identify
the relevant good to bad and bad to good transition that
involve only two good parkers. These are illustrated in
Figs. 4b and Fig. 4c.
As these events are expected to dominate the dynamics, one
can write the following equation for the evolution of $\rho$
once the jamming limit has been reached:
\begin{equation}
d\rho/dt = R_{bg} - R_{gb} + h.o.t..
\end{equation}
Here $R_{bg}$ and $R_{gb}$ are the rates of bad to good and good
to bad transitions respectively, and $h.o.t$ are collective
transitions that a larger number of particles, and hence,
occur at a slower rate.

\begin{figure}[htb]
\epsfxsize = 3.2in
\centerline{\epsffile{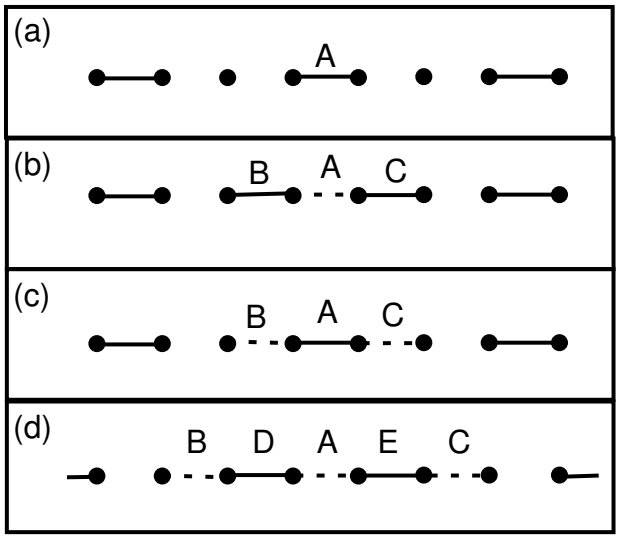}}
\caption{ (a) This illustrates the concept of a ``bad'' parker.
The bond at location A is the bad parker,
as it leaves two sites free. (b) This illustrates a transition
from one bad parker to two good parkers. The dashed line represents
the original bad parker at location A that desorbs. Then, the
two good parkers, B and C, adsorb. (c) This illustrates the
two good parkers to one bad parker transition. In this case,
the two parkers at B and C leave and one attaches at A.
(d) This illustrates a spatial arrangement that corresponds
to a two-to-three particle transition. The two possible bond
distributions are given by the dashed lines at sites
A, B, and C (the good parkers) and the solid lines at
sites D and E, respectively.}
\end{figure}

We were able to track the bad to good and good to bad
transitions during the simulation. This was accomplished by
converting the particle sites to an array of bond locations.
Each location between two
sites was assigned a value of 1 if a bond was present and 0 is
there was no bond. For example, the solid lines in Fig.~4b would
be represented by the string $1 0 1 0 1 0 1$. Notice, by definition,
between any two bonds there is an open space, so the completely
filled system is represented by $1 0 1 0 1 0 1 0 1 0 \ldots$.
The string of bond locations was saved at step $i$ and $i+\Delta$.
Each bond location was taken as the initial digit in a seven digit
string, and these strings were compared for steps $i$ and
$i+\Delta$. We counted the following transitions:
\begin{eqnarray*}
1 0 1 0 1 0 1 & \Longleftrightarrow & 1 0 0 1 0 0 1.
\end{eqnarray*}
These transitions correspond to two good to one bad and one
bad to two good, as discussed in Fig.~4c and Fig.~4b respectively.
The choice of $\Delta$ is important.
If $\Delta$ is too small, the transitions do not have enough
time to complete. For example, in the extreme limit of choosing
$\Delta$ to be a single time step, it is not possible to have
multiparticle events, but the total number of bound sites can
change by one. Essentially, $\Delta$ must be large enough
for the multiparticle transitions to have time to complete.
For $\Delta$ large enough, the recorded
number of transitions is essentially independent
of $\Delta$. For the data reported here,
we used $\Delta = 2 \times 10^6$.

In addition to counting multiparticle transition, we also
recorded the total change in $\rho$. Figure 5 compares
the actual value of $\rho$ as a function of the number of iterations
with the value obtained using
Eq. 2 and the computed number of bad to good and good to bad
transitions. Once the jamming limit
is reached, the bad to good and good to bad
transitions account for 94.3\% of the change in $\rho$,
confirming the general idea behind Eq. 2.
An additional 3.2\% of the change in $\rho$
is accounted for by considering a single class of three particle
transitions
where three good parkers were replaced by two parkers, and
the reverse process. These were counted by considering 9 digit
strings and looking for the transition:
\begin{eqnarray*}
1 0 1 0 1 0 1 0 1 & \Longleftrightarrow & 1 0 0 1 0 1 0 0 1.
\end{eqnarray*}
This curve is also plotted in Fig. 5.
The spatial arrangement corresponding to
this transition is shown in Fig.~4d. It is important to note
that when bonds exist at site D and E, the only way to increase
the number of bound sites in this region is for two particles
to desorb and three particles adsorb at sites A, B and C.

\begin{figure}[htb]
\epsfxsize = 3.2in
\centerline{\epsffile{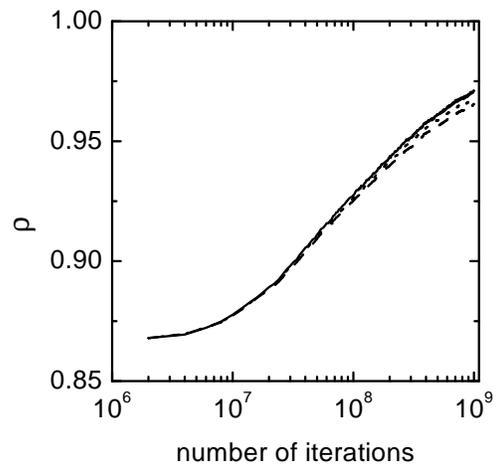}}
\caption{Shown here is the coverage fraction $\rho$ as a function of
the number of iterations for the one-dimensional model and
$K = 1000$ (solid curve). The curve is plotted starting at the end of
the jamming limit plateau. Also plotted are
two curves that are obtained by numerically integrating Eq. 2
using the coverage fraction at the jamming limit as the initial state.
The dashed line is the result when only the rates for the good
to bad and bad to good transitions are included in Eq. 2.
These transitions are described in Fig. 4. The dotted line shows
the improvement at late times by including a single higher-order
transition involving three good parkers converting to two parkers.}
\end{figure}

\section{Two-dimensional Simulations}

The results of the simulation for $\rho$ as a function of iteration
step for the adsorption of dimers on a
two-dimensional triangular lattice (Model A)
and on a two-dimensional triangular lattice with nearest-neighbor
exclusion
(Model B) are presented in Figs. 6 and 7, respectively. For Model A,
previous simulations have found a
jamming limit of 0.9243 \cite{BK97}. Our simulations give a value of
0.9120. For Model B, we find a jamming limit
of 0.275. Recall that complete coverage in this case corresponds to
$\rho = 0.4$. We are not aware of any previous work on a Model B type
simulation. However, by appropriately including the empty
nearest-neighbor sites in the definition of
$\rho$, we can compare to simulations involving n-mers of length 6 that
cover a hexagonal patch. These simulations find a jamming limit
of 0.6847, and our converted value is 0.6875 \cite{BK97}.

\begin{figure}[htb]
\epsfxsize = 3.2in
\centerline{\epsffile{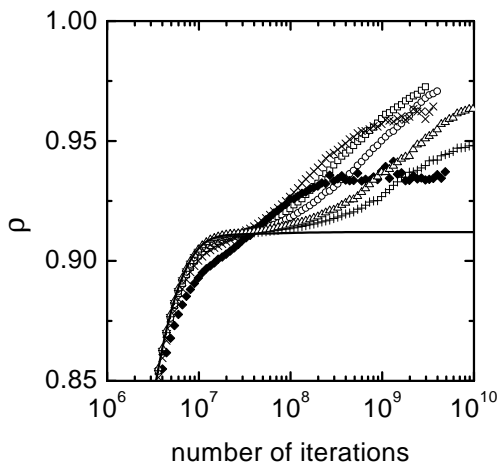}}
\caption{Shown here is the coverage fraction $\rho$ as a function of
the number of iterations for Model A. The solid
line represents the evolution for $K^{-} = 0$. The other curves are
for: solid diamonds ($K = 200$), crosses ($K = 500$),
open squares ($K = 1000$), open circles ($K = 2000$),
open triangles ($K = 5000$), and plus signs ($K = 10000$).}
\end{figure}

The results for the two-dimensional cases are qualitatively similar to
the one-dimensional case. One observes multiple
time scales: a rapid approach
to the jamming limit and a slow relaxation to the steady-state value.
This suggests that the same picture of multiparticle transitions
will apply to the two-dimensional system.
However, in contrast to the one-dimensional case, the identification
of collective transitions is significantly more complex
for Model A and B because of the number of arrangements due to
differing orientations of the bonds that can
produce bad parkers. However, we did carry out a limited analysis
for the case of Model A.

\begin{figure}[htb]
\epsfxsize = 3.2in
\centerline{\epsffile{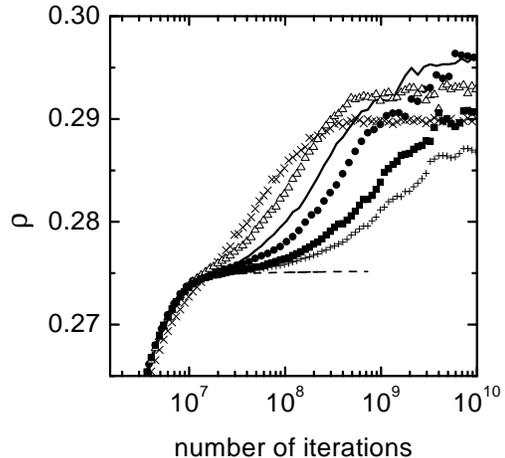}}
\caption{Shown here is the coverage fraction $\rho$ as a function of
the number of iterations for Model B. The dashed
line represents the evolution for $K^{-} = 0$. The other curves are
for: crosses ($K = 200$), open triangles ($K = 500$),
solid line ($K = 1000$), closed circles ($K = 2000$),
closed squares ($K = 5000$), and plus signs ($K = 10000$).}
\end{figure}

The method used to track multiparticle events in Model A was
similar in concept to the one dimensional case. However, because
the bonds have orientation, we compared the actual sites
instead of the bonds for three classes of transitions:
\begin{eqnarray*}
0110 & \Longleftrightarrow & 1111 \\
& & \\
{0\;\;1\;\; \atop \;\;1\;\;0} & \Longleftrightarrow &
{1\;\;1\;\; \atop \;\;1\;\;1} \\
& & \\
{\;\;1\;\; 0 \atop 0 \;\;1\;\;} & \Longleftrightarrow &
{\;\;1 \;\;1 \atop 1\;\; 1\;\;}
\end{eqnarray*}
In this case, occupied sites are represented by 1 and unoccupied
sites are represented by 0. Using sites instead of bonds results
in some differences between the methods used in
the two-dimensional and one-dimensional
cases. First, the transitions counted in this
manner correspond to classes of transitions in the following sense.
Because we track sites and not bonds, two neareset neighbor sites
can be occupied either because they share a bond or because of two
neighboring bonds that are at an angle to the line being considered.
So, the first class of transitions includes the transitions that
are exactly analogous to the one-dimensional good to bad transitions.
But, it also includes multiparticle transitions that involve
bonds at an angle to the horizontal and that successfully fill
the empty sites along the horizontal. Second, the offset of
the 1's and 0's in the second two classes of transitions are
important and reflect the underlying hexagonal lattice. Note
that because only nearest-neighbor bonding is allowed, the
diagonal connecting the two zeros in each case is not an allowed
binding site. Finally, for the two-dimensional case,
we exploited the hexagonal symmetry of the problem, and
multiplied the rate for the first type of transition
by three. The rates of the second two transitions are multiplied
by 3/2 to account for double counting. The value of $\Delta$
was chosen in a similar fashion to the one-dimensional case
and corresponded to a constant interval in $\Delta \rho = 0.005$.

The results for $\rho$ as a function of the number
of iterations and the value of $\rho$ computed from Eq. 2 using just
these three classes of events defined above
are plotted in Fig. 8. One striking
feature of Fig. 8 is the fact that the two-particle
events we identified account for nearly 100\% of the dynamics
until the number of steps reaches approximately $2 \times 10^9$.
At this point, the coverage
continues to grow, only there is essentially no change due to the
identified two-particle events. This strongly suggests other
two-particle events or higher-order
events involving more than two particles are becoming important.

\begin{figure}[htb]
\epsfxsize = 3.2in
\centerline{\epsffile{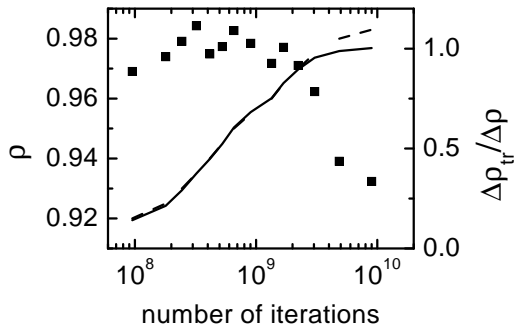}}
\caption{In this figure, the curves are the coverage fraction
$\rho$ versus number of iterations for Model A. A reduced
grid of 600 by 600 was used to facilitate the counting
of collective events. The dashed line
is the result of the simulation for $K = 5000$. The solid
line is the result obtained by tracking two-particle events and
integrating Eq. 2 starting at the jamming limit.
The symbols are plotted against the right-hand
axis and give the ratio of the change in $\rho$ as computed by
the two methods. The agreement between the two methods
is excellent until approximately $2 \times 10^{9}$ steps. At this
point, the contribution to the dynamics of two-particle events
decreases dramatically.}
\end{figure}

\section{Possible Application to Langmuir Monolayers}

An obvious question is: do the triangular lattices considered here apply
to any experimental systems? There is indirect evidence that the models
discussed here are relevant to the binding of Ca$^{++}$ ions to
a Langmuir monolayer. Langmuir monolayers are composed of insoluble,
amphiphilic molecules that are confined to the air-water
interface \cite{Mono}. They exhibit the usual gas, liquid, and
solid phases, as well as a large number of two-dimensional
analogs of smectic phases \cite{Phase}.
Many of these phases are hexatic, with
the molecules locally arranged on a distorted hexagonal lattice.
When Ca$^{++}$ is present in the water, it can bind two 
fatty-acid molecules together. This substantially alters a number
of the physical properties of
the monolayer, such as the lattice spacing and the
viscosity \cite{SBMZD92}.
Existing measurements \cite{KTO88} and models \cite{AF91} of Ca$^{++}$
binding have focused on
the equilibrium coverage fraction. However, the
measurements have focused on time scales of one hour or less.
The coverage fraction depends strongly on pH, which is
understandable in terms of the
degree of ionization of the fatty acid headgroup. At low values
of the pH, essentially all of the fatty acid molecules are neutral,
and the Ca$^{++}$ ions do not bind. As the pH is increased, an
increasing number of fatty acid molecules become charged, and the
Ca$^{++}$ ions are free to bind to the monolayer.

The possible relevance of Model A and B to the fatty
acid monolayers
is based on viscosity measurements
as a function of time in the presence of Ca$^{++}$ for
the hexatic phase of a particular fatty acid \cite{GCD99}.
Figure 9 reproduces one set of data from Fig.~2
of Ref.~\cite{GCD99}, illustrating
a typical time evolution of the viscosity. 
The viscosity increases 2 orders of magnitude
over 15 hours. The time evolution can be divided into three distinct
regions: an initial rapid rise
in viscosity within the first hour, a slower rise in viscosity
covering 5 - 6 hours, and a final even slower rise in viscosity.
For comparison, the computed fractional coverage of $\rho$ is
shown in Fig. 9 versus the number of iterations. In this case,
we have used a linear scale for the number of iterations. The
previous plots all used a logarithmic scale.
The time evolution of the Ca$^{++}$ binding exhibits the
three general regions present in the viscosity data, and as such,
provides a natural explanation for the effect.

There are a number of points with regard to the connections
between the model and the monolayer experiments. The
simulations are consistent with the fact that previous measurements
of Ca$^{++}$ binding do not observe multiple time scales. In the
simulations, the interesting change
in coverage fraction occurs at late times,
while in the experiments, only relatively early times are
considered \cite{KTO88}. Also, the fact that the experiments agree
reasonably well with equilibrium calculations \cite{AF91} is not
surprising because the late-time changes in $\rho$ are
relatively small in the simulations.
Therefore, longer experiments with more precise measurements
of $\rho$ are required to directly observe the effects predicted
by our simulations. This discussion naturally leads
to the second point: how do small changes in coverage fraction
produce large
changes in viscosity? An ad hoc model that is capable of
explaining the large viscosity rise assumes that 
the viscosity is proportional to 1/(A - $\rho$), where
A is a constant determined by the equilibrium coverage fraction.
This model is based on the idea that the fluidity (the inverse
of the viscosity) is proportional to the number of
unbound sites. Clearly, both more careful direct measurements
of the coverage fraction versus time and a better theoretical
understanding of the connection between viscosity and
coverage fraction are needed.

\begin{figure}[htb]
\epsfxsize = 3.2in
\centerline{\epsffile{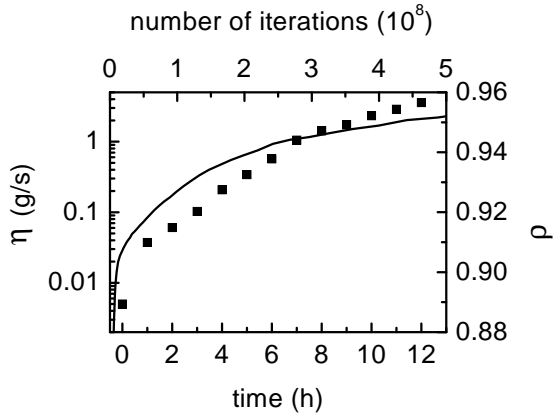}}
\caption{In this figure, the solid points are taken from Fig.~2
in Ref.~\protect\cite{GCD99}. They are the viscosity values (left-hand
axis) versus time (bottom axis) for pH = 5.5
and a Ca$^{++}$ concentration in the subphase of 0.65 mM.  The
solid curve is the simulation data for Model A and $K = 500$.
Plotted here is coverage fraction $\rho$ (right-hand axis)
versus number of iterations (top axis).}
\end{figure}

The final two comments concern possible refinements of
the adsorption model when applying it to the monolayer system.
In this paper, we considered the two cases of binding to any
open pair of sites (Model A) and binding with nearest-neighbor
exclusions (Model B) because they are simple cases with
different geometric arrangements. The correct detailed description
of the Langmuir monolayer system is certainly more complicated
than either of these. However, as mentioned, the degree of ionization
of the monolayer is pH dependent. To zeroth order, Model B is a
reasonable description of a monolayer that is
only partially ionized for two reasons.
First, for a partially ionized monolayer, if
a particular site is available for binding, it is highly unlikely
that any of the neighbors will be available as well. Second,
the steady-state values of $\rho$ found in Model B are in reasonable
agreement with measurements of the values of $\rho$ reported
for monolayers for pH between 5 and 6 \cite{AF91}.

The second refinement concerns lateral diffusion of Ca$^{++}$
ions once they have bound to the monolayer. Inclusion of
diffusion should not substantially alter the qualitative
results presented here, but it would effect the quantitative
interpretation of the rate constants $K^+$ and $K^-$. One can model
lateral diffusion of Ca$^{++}$ as the unbinding of a Ca$^{++}$
from one of the monolayer molecules followed by
a rotation around its remaining bond and subsequent
binding to another available site. However, this process could
also be viewed as a complete unbinding and rebinding at a neighboring
site with a renormalized rate constant. In addition, for $\rho$ to
evolve in time, diffusion would need to be coupled with additional
binding. This would result in
rearrangements that are completely analogous to transitions from
one bad parker to two good parkers. Therefore, even with diffusion
in the plane of the monolayer, the basic physics remains the same.
Jamming will still occur, and the slow relaxation of the bad parkers
due to cooperative behavior will result in the slow time scales. 

\section{Discussion}

Equation 2 provides a means of expanding the dynamics in terms
of collective events that occur on slower and slower
time scales. We were able to directly confirm this in the
simple situation of the one-dimensional model and for the
two-particle transitions in Model A. For the one-dimensional
case, two-particle events were sufficient to describe the
dynamics of the system, as was found in the continuous
model. This results in two plateaus in the time evolution
of the coverage fraction. Single particle events,
dominated by adsorption, rapidly drive the system to the
jamming limit. Processes involving two particles are sufficiently
slow that $\rho$ plateaus for some time. The length of this
plateau is controlled by $K$, as $K$ ultimately determines
the rates of multiparticle transitions. The larger the
value of $K$, the longer the system remains at the jamming limit.
After enough time, the two particle processes have
a sufficiently large contribution
to the dynamics that $\rho$ increases at
a noticeable rate until the true steady-state value is reached,
and the coverage plateaus again.

In contrast, one can imagine more complicated dynamics,
such as multiple plateaus in the time evolution,
occurring when collective events involving
three or more particles are important. For example,
Fig.~4d illustrates the existence of spatial arrangements of
unbound sites that can not be corrected by two-particle
events. In Model A, Fig.~8 shows that
the two particle events are not capable of
bringing the system to its steady-state value, as they
are no longer contributing to the dynamics at late
enough times. This suggests that the remaining unbound
sites occur in spatial arrangements that are analogous
to those in Fig.~4d. Multiple
plateaus would arise in the
extreme case where the transition rates for two-particle
and three-particle events are sufficiently
different. This would occur as follows. The
two-particle transitions would drive the system to
some value $\rho_2$ in a given time $t$. If $t$ was small
enough compared to the three-particle transition rate,
the system would stay at $\rho_2$ until the three-particle
events contributed to the dynamics. 

Identifying the existence of multiple plateaus is
extremely challenging. First, the steady state value
of $\rho$ must be sufficiently large that at late
times the unbound sites are arranged in such a way
that two-particle events are ineffective. This implies
a sufficiently high value of $K$. However, this in
turn both decreases significantly the rate of collective
events and increases the time to reach steady state.
For Model A and B, we have indirect evidence of
multiple plateaus. In both cases,
the coverage fraction for $K = 10000$ appears to be
leveling off at a value that is lower than the apparent
steady-state values for
$K = 500$, in the case of Model A, and $K = 200$, in the
case of Model B. In principle, $\rho(\infty)$ should
approach one (or 0.4 for Model B) as $K$ approaches
infinity. Therefore, the behavior for $K = 10000$
suggests the beginning of a secondary plateau.
Unfortunately, as discussed, the time required to achieve
steady-state increases with $K$, and
we do not have sufficient computing power to
determine if this is a true intermediate plateau
for $K = 10000$ or if this is actually the steady-state
value.

It is clear that both analytic and more numeric
work is needed to fully explore the effects of
higher order transitions. 
Identification of the higher order
terms in Eq. 2 is an important step in
this process. An exhaustive identification
of all possible transitions is beyond the scope of this paper;
however, Fig.~10 identifies a small subset of transitions
that illustrates why one would expect differences
between Model A and B for large enough times or large
enough values of $K$.

Figure 10a  shows a set of
transitions for Model A, and
Fig. 10b illustrates the equivalent ones
for Model B. In both cases,
there exists at least two different
classes of transitions that turn two good parkers
(labeled A, B, and C in Fig. 10)
into one bad parker. For Model A,
if A and C desorb, then there are two possible sites that result in a
bad parker, and two possible sites that result in the re-establishment
of a pair of good parkers.  But, if A and B desorb, then the situation
reverts to the one-dimensional case. (In one-dimension, after two good
parkers desorb, bonding to one out of the three open sites corresponds
to the creation of a bad parker (see Fig. 4c).)  
For Model B, if A and C desorb, then there are six
possible sites that result in a bad parker, and six possible sites that
result in the re-establishment of a pair of good parkers. This results
in the same
probabilities as in Model A.
However, in the A to B case,
two sites are available for bad parkers, and two sites are
available for good
parkers. Therefore, the chance of two good switching to one bad is
increased. Because differences in transition rates
may affect the length of any additional plateaus, detailed
calculations of these rates are needed for a fuller understanding
of the possible dynamics.

\begin{figure}[htb]
\epsfxsize = 3.2in
\centerline{\epsffile{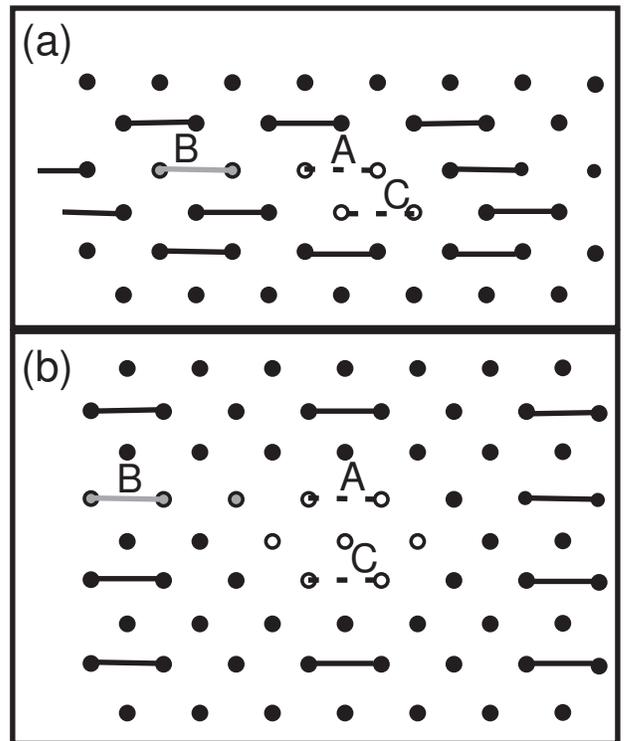}}
\caption{This figure illustrates collective events in the two-dimensional
models. (a) Initially, there are particles at the locations labeled A, B,
and C. There are two possibilities. If A and C desorb, the open circles
represent the now available sites. If A and B desorb, the open circles
at A and the gray circles are now the available bonding sites. (b)
Again, particles are initially at the location A, B, and C. If A and C
desorb, the open circles represent the now available sites. If A and
B desorb, the open circles at A and the gray circles are now the
available bonding sites. In this case, there is an additional available
site because of the nearest-neighbor exclusion.}
\end{figure}

In conclusion, we present results of simulations of the
reversible parking lot model for three different lattices.
We have directly confirmed the importance of multiparticle
transitions for governing the late time behavior in two
of the models. The behavior of the third model is consistent
with the other two. We discussed the implications of a
description of the dynamics in terms
of collective events. For the right ratios of transition
rates, one would expect to observe multiple plateaus.
There is a suggestion of intermediate plateaus in our system,
but computational limits prevented any conclusive evidence.
One alternative method for finding multiple plateaus would
be to consider different particle shapes as a means of
adjusting the
relative rates of multiparticle transitions.
Finally, we presented the possible relevance of the model
to the binding of Ca$^{++}$ to Langmuir monolayers. We
showed that the jamming and subsequent slow relaxation of the
binding of Ca$^{++}$ ions is a strong candidate for
the source of the long-time scales observed in the viscosity
measurements. There are experimental and
theoretical details that require further exploration, including
direct measurements of the Ca$^{++}$ coverage fraction,
modeling of the dependence of viscosity on Ca$^{++}$
coverage fraction, better modeling of pH effects, and
both measurements and modeling of lateral diffusion.
However, given how well the model presented here captures
the time scales present in the viscosity data,
such future studies should prove extremely fruitful.

\acknowledgments

We thank Amy Kolan for bringing the parking lot model to
our attention and fruitful discussions with Chuck Knobler and
Robijn Bruinsma. This work was supported in part
by NSF grant CTS-9874701. Acknowledgment by M. Dennin is made to
the donors of The Petroleum Research Fund, administered by the
ACS, for partial support of this research.


\begin{references}

\bibitem{E93} A large literature has developed in this field. 
For a review,
see J. V. Evans, Rev. Mod. Phys. {\bf 65}, 1281 (1993).

\bibitem{SVS98} For a review, see P. Schaaf, J. C. Voegel, and
B. Senger, Ann Phys. Fr. {\bf 23}, 1 (1998).

\bibitem{TST90} G. Tarjus, P. Schaaf, and J. Talbot, J. Chem. Phys.
{\bf 93}, 8352 (1990).

\bibitem{J94} X. Jin, Ph.D. Thesis, Purdue University, West
Lafayette, 1994.

\bibitem{JTW94} X. Jin, J. Talbot, and N. H. Linda Wang, AIChE J.
{\bf 40}, 1685 (1994).

\bibitem{RL87} {\it Non-Debye Relaxation in Condensed Matter}, edited by
T. V. Ramakishnan and M. Raj Lakshmi (World Scientific, Singapore, 1987).

\bibitem{MPV87} M. Mezard, G. Parizi, and M. A. Virasoro, {\it Spin
Glass Theory and Beyond} (World Scientific, Singapore, 1987).

\bibitem{WSB96} {\it Traffic and Granular Flow}, edited by D. E. Wolf,
M. Shreckenberg, and A. Bachem (World Scientific, Singapore, 1996).

\bibitem{R58} A. Renyi, Publ. Mat. Inst. Hung. Acad. Sci. {\bf 3},
109 (1958).

\bibitem{JTT94} X. Jin, G. Tarjus, and J. Talbot, J. Phys. A {\bf 27},
L195 (1994).

\bibitem{PB97} V. Privman and M. Barma, J. Chem. Phys. {\bf 97},
6714 (1992).

\bibitem{KB94} P. L. Krapivsky and E. Ben-Naim, J. Chem. Phys.
{\bf 100}, 6778 (1994).

\bibitem{NKBJN98} E. R. Nowak, J. B. Knight, E. Ben-Naim, H. M. Jaeger,
and S. R. Nagel, Phys. Rev. E {\bf 57}, 1971 (1998).

\bibitem{KNT99} A. J. Kolan, E. R. Nowak, and A. V. Tkachenko,
Phys. Rev. E
{\bf 59}, 3094 (1999).

\bibitem{F39} P. J. Flory, J. Am. Chem. Soc. {\bf 61}, 1519 (1939).

\bibitem{CR63} E. R. Cohen and H. Reiss, J. Chem. Phys. {\bf 38},
680 (1963).

\bibitem{W66} B. Widom, J. Chem. Phys. {\bf 44}, 3888 (1966).

\bibitem{BK97} Lj. Budinski-Petkovi\'{c} and U. Kozmidis-Luburi\'{c},
Phys. Rev. E {\bf 56}, 6904 (1997).

\bibitem{Mono} For reviews of Langmuir Monolayers, see H. Mohwald, 
Annu. Rev.
Phys. Chem. {\bf 41}, 441 (1990); H. M. McConnell, {\it ibid.} {\bf 42},
171 (1991).

\bibitem{Phase} For a review of phase transitions in monolayers, see C. M.
Knobler and C. Desai, Annu. Rev. Phys. Chem. {\bf 43}, 207 (1992).

\bibitem{SBMZD92} M. C. Shih, T. M. Bohanon, J. M. Mikrut, P. Zschack,
and P. Dutta, J. Chem. Phys. {\bf 96}, 1556 (1992).

\bibitem{KTO88} K. Kobayashi, K. Takaoka, and S. Ochiai, Thin Solid
Films {\bf 159}, 267 (1988).

\bibitem{AF91} D. J. Ahn and E. I. Franses, J. Chem. Phys. {\bf 95},
8486 (1991).

\bibitem{GCD99} R. S. Ghaskadvi, S. Carr, and M. Dennin,
J. Chem. Phys. {\bf 111}, 3675 (1999).


\end{references}
\end{document}